\documentstyle[aps,prl,multicol,epsf,graphics]{revtex}

\title{Reply to "Comment on 'A linear optics implementation of weak values in Hardy's paradox'"}

\author{S. E. Ahnert and M. C. Payne}
\address{Theory of Condensed Matter Group, Cavendish Laboratory, \\
Madingley Road, Cambridge CB3 0HE, U.K.}

\begin{document}
\maketitle 
\begin{abstract}

The comment by Lundeen et al. contains two criticisms of our proposal. While we agree that the state-preparation procedure could be replaced by a simpler setup as proposed by the authors of the comment, we do not agree with the authors on their second, and more important point regarding two-particle weak measurements. We believe this to be the result of a misunderstanding of our original paper \cite{us}.

\bigskip

\end{abstract}

\begin{multicols}{2}

Our understanding of weak measurement is based primarily on the analysis of Duck {\em et al.} \cite{Duck} which clarified the concepts presented in the original 1988 paper by Aharonov {\em et al.} (AAV) \cite{Aharonov}. What is called 'weak measurement' can be understood as a three-stage process:

\begin{itemize}
\item[I)]{State preparation of a finite dimensional quantum system}
\item[II)]{Entangling of this system with the state of another quantum system. Conventionally (but not necessarily) a continuous variable is chosen. This quantum system (e.g. a Gaussian beam or wavepacket) plays the role of the measurement pointer.}
\item[III)]{Post-selection of a particular state in the Hilbert space of the system in I, projecting the system in II into a superposition with complex coefficients.}
\end{itemize}

The important characteristic of weak measurement is that the system in II - the pointer state - is, through post-selection in III, projected into a superposition with complex coefficients. This projection can give rise to a highly non-trivial distribution of possible pointer positions. 

The only requirement in terms of 'weakness' of this measurement is that the possible states in II - corresponding to the different eigenstates of I - have an overlap, so that the projection in III can produce interesting, non-trivial results. 

We think that Lundeen {\em et al.} have misunderstood equation (23) of our paper. A large part of their argument is based on showing that the operator of equation (18) is separable. However our paper does not claim anything to the contrary, or indeed that the outcome of (23) corresponds to any of the four joint operators $|HH\rangle \langle HH|$, $|HV\rangle \langle HV|$, $|VH\rangle \langle VH|$, $|VV\rangle \langle VV|$ in \cite{Popescu}. As we say in the paragraph following equation (23):

\bigskip
{\em [...] Eq. (23) reveals in a very straightforward way that the three weak values $(\gamma, \epsilon)$, $(\epsilon, \gamma)$ and -$(\gamma, \gamma)$ - which correspond directly to the values in Eq. (13) - combine to give the paradoxical result $(\epsilon, \epsilon)$. It is paradoxical because $(\epsilon, \epsilon)$ is the result corresponding to the combined polarization state $|V_2V_4\rangle$, which was projected out. [...]}

\bigskip

The values in Eq. (13) are the joint particle occupation operators 1, 1, -1, and 0, which are mirrored in terms appearing in the sum of $(\gamma, \epsilon)$, $(\epsilon, \gamma)$, $-(\gamma, \gamma)$ and the nonexistent $(\epsilon, \epsilon)$. We regard the joint occurrence of arrival time pairs, such as $(\gamma, \epsilon)$ to be equivalent to one single two-particle pointer result, telling us something about where the particles are.

The whole point however is that the outcome of Eq. (23), namely $(\epsilon, \epsilon)$ is the result of the sum of these values. It is paradoxical, as the arrival time $\epsilon$ implies the vertical polarization, but the state $|V_2V_4\rangle$ has been projected out.

In terms of our description of weak measurement this means that since the measurement pointer is a quantum object, we have produced a superposition of measurement pointer positions which - because it contains destructive interference - points to somewhere it should not be allowed to point in a conventional measurement - just like the "spin value of a 100" in Aharonov's original 1988 paper \cite{Aharonov}. Eq. (23) demonstrates exactly this superposition. 

Thus $(\epsilon, \epsilon)$ is not, as Lundeen {\em et al.} claim "in contradiction with Aharonov's prediction" of occupation number 0 for the VV case, since it is not a joint measurement of the photon polarization. It is a demonstration of why Aharonov's values work - they form a superposition of joint measurement values, which gives rise to the paradoxical combined single particle result $(\epsilon, \epsilon)$. After all it is only in the combination of the two single particle results that the paradox arises. 

Concerning the operators designed to measure the individual joint particle occupation numbers, which are mentioned in the penultimate paragraph of our paper, Lundeen {\em et al.}'s criticism is that these measurements are strong, not weak. As clarified above in our description of weak measurement, this process does not have to involve a 'weak' interaction. What is required is an overlap of the states corresponding to different measurement results. This overlap exists in the case of the measurement of equation (23). When measuring individual joint particle occupation operators, there cannot be any overlap of pointer states, since we are measuring eigenstates of the joint state of photon polarization. The operators hence cannot give anything but the results $(\epsilon, \gamma)$, $(\gamma, \epsilon)$ and -$(\gamma, \gamma)$, where the phase of -1 of the last term is only detectable by performing the full measurement of Eq. (23). To speak about these individual measurements in terms of weak or strong measurements does not make sense: We can have no superposition of pointer positions, as we are measuring eigenstates of the system.

For example, in the notation of \cite{Popescu}, the pointer could be centred on the $|NO\rangle \langle NO|$ position (corresponding to the non-overlapping arms in Hardy's paradox) which is equivalent to the $|HH\rangle \langle HH|$ eigenstate in our setup. But since there would be no other pointer functions to interfere with, the overall phase factor of -1 (which corresponds the value of the weak two particle occupation number $N^{+,-}_{NO,NOw} = -1$) of this particular pointer wave function cannot be determined.

One can measure the presence or absence of pointer terms in the superposition (a rather trivial knowledge which is also given if one knows the initial state), but the complex factors, which arise because of the weak measurement and which create the non-intuitive results in all weak measurements, cannot be measured outside a superposition. 

Furthermore, contrary to what Lundeen {\em et al.} state, the weak values do appear in this setup, namely in eq. (23), where they form a superposition. The values 0, 1, 1 and -1 can be deduced from the overall result $(\epsilon, \epsilon)$ of this equation in combination with the moduli of the coefficients for $(\epsilon, \gamma)$, $(\gamma, \epsilon)$ and $(\gamma, \gamma)$. As discussed above these moduli (indicating the presence or absence of pointer terms) can be measured too, but in this case are known because the initial state is known. 

Regarding the state-preparation method, Lundeen {\em et al.} do have a point. It is true that the setup shown in Fig. 3 of the original paper \cite{us} does not erase the which-path information of the photons emerging from the vertical ports of the two PBSs left of the sources. However this information could in principle be erased, and the implication was that these outputs are never measured. Admittedly it would be easier (and experimentally much more feasible) to choose a different, simpler state-preparation method as suggested by Lundeen {\em et al.}. 

Finally, the claim that there already was a clear way to measure weak values in Hardy's paradox using linear optics is incorrect: The paper referred to in the comment which implements Hardy's paradox with linear optics \cite{Bouw} (without a mention of weak values) was unpublished when our paper appeared. The other paper which the authors of the comment refer to in this context \cite{Resch} might provide a method which is "ideally suited for linear optics", but does not provide an explicit experimental proposal.

\end{multicols}
\end{document}